\documentstyle[epsfig]{mn2e}

\def\simgt{\mathrel{\lower0.6ex\hbox{$\buildrel {\textstyle >}
 \over {\scriptstyle \sim}$}}}
\def\simlt{\mathrel{\lower0.6ex\hbox{$\buildrel {\textstyle <}
 \over {\scriptstyle \sim}$}}}

\newcommand{\Msolar}{\mbox{\,$\rm M_{\odot}$}}        

\newcommand{\gtsim}{\mbox{{\raisebox{-0.4ex}{$\stackrel{>}{{\scriptstyle\sim}}
$}}}}

\begin{document}

\title[The discovery of a type-II quasar at $z = 1.65$ with integral-field spectroscopy]{The discovery of a type-II quasar at $z = 1.65$ with integral-field spectroscopy\thanks{Based on observations performed at the European Southern Observatory, Chile [programme ID: 71.B-3015(A)].}}

\author[M.~J.~Jarvis et al.]
{Matt J.~Jarvis$^{1}$\thanks{Email: m.jarvis1@physics.ox.ac.uk}, Caroline van
  Breukelen$^{1}$ \& R.J.~Wilman$^{2}$ \\
\footnotesize\\
$^{1}$Astrophysics, Department of Physics, Keble Road, Oxford, OX1 3RH. \\
$^{2}$Department of Physics, University of Durham, Durham DH1 3LE
}
\maketitle

\begin{abstract}
In this letter we report the serendipitous discovery of a genuine
type-II quasar at $z = 1.65$ using integral-field data from VIMOS on the VLT. This
is the first discovery of a type-II quasar at $z > 1$ from optical data
alone. J094531-242831, hereafter J0945-2428, exhibits strong narrow
($v < 1500$~km~s$^{-1}$) emission lines, has a resolved host galaxy,
and is undetected to a radio flux-density limit of $S_{5 \rm GHz} =
0.15$~mJy ($3\sigma$). All of these lead us to believe that J0945-2428 is a
bona fide type-II quasar.

The luminosity of the narrow-emission lines in this object suggest
that the intrinsic power of the central engine is similar to that
found in powerful radio galaxies, indicative of similarly large
supermassive black hole of $\sim 3 \times 10^{8}$~M$_{\odot}$
(assuming that it is accreting at its Eddington limit). However, from
near-infrared imaging observations we find that the old stellar
population in the host galaxy has a luminosity of $\sim
0.2~L^{\star}$, mildly inconsistent with the correlation between
black-hole mass and bulge luminosity found locally, although the
uncertainty in the black-hole mass estimate is large.

This discovery highlights the power that integral-field units have in
discovering hidden populations of objects, particularly the sought
after type-II quasars which are invoked to explain the hard X-ray
background. As such, future large integral-field surveys could open up
a new window on the obscured accretion activity in the Universe.

\end{abstract}

\section{INTRODUCTION}

Under the unification picture of active galactic nuclei (AGN), radio galaxies
and radio-loud quasars are believed to be the same type of
object. Under this unification scheme only viewing angle dictates
whether we see the object optically as a quasar with broad permitted
lines in emission and an unresolved optical core, or as a radio galaxy
with only narrow emission lines and where the host galaxy is resolved
(see e.g. Urry \& Padovani 1995). 

Applying the same reasoning to radio-quiet quasars, which outnumber
their radio-loud counterparts by a factor of $\sim 10-100$
(e.g. Goldschmidt et al. 1999), there should be a population of
radio-quiet quasars which are obscured in the same way as radio
galaxies such that their quasar nuclei shine along the plane of the
sky, the so-called type-II quasars. However, such sources are extremely difficult to detect, because
they would appear as normal elliptical galaxies in optical or near-infrared images and only spectroscopy would detect the strong narrow-emission lines.

From an X-ray perspective the hypothesis that there must be a
population of type-II quasars, in which the direct light from the quasar
is obscured and only narrow emission lines are observed, is a
cornerstone assumption for models of the hard-X-ray background
(XRB; e.g. Wilman \& Fabian 1999; Wilman, Fabian \& Nulsen 2000). This requirement for type-II quasars by models of the XRB fits in
well with the orientation based arguments for radio-loud objects.

With the advent of the latest generation of X-ray satellites (CHANDRA
and XMM-Newton), the 
majority of the XRB in the
Chandra band ($< 8$~keV) has been resolved (e.g. Alexander et al. 2003). The bulk of which is
produced by Seyferts and quasars with moderate obscuration
(N$_{\rm H} < 2 \times 10^{23}$~cm$^{-2}$), where the Seyferts evolve
quite rapidly to $z=0.8$ and 
the quasars evolve somewhat more slowly. But of this population, only a handful
are bona fide type-II quasars (e.g. Gandhi et al. 2004).

The situation at harder energies is quite different. The majority of
the hard sources lie at the faint end of the flux distribution
(e.g. Fig.~2 in Giacconi et al. 2001) and only 50 per cent of the XRB
above 7 keV is resolved (e.g. Worsley et al. 2004). The unresolved flux has the spectrum of a
highly obscured AGN,
suggesting that the range of column density at $z \sim 1$ is similar
to that locally, implying that there is a bias towards highly
obscured, Compton-thick objects. This essentially means that there may
be many obscured AGN which have yet to be discovered in the large,
well-studied fields in the literature.

In addition to the high obscuring column densities causing
the hardening of the X-ray spectra there are other reasons why some
quasars may be obscured and others not. These include factors such as
intrinsic luminosity (e.g. Simpson, Rawlings \& Lacy 1999; Ueda et
al. 2003; Hasinger et al. 2004) and the rate at which the black-hole
accretes mass (e.g. Fabian 1999), which may be needed to derive a
complete unification model for active galactic nuclei (AGN). This is
borne out by the diversity of the hard-X-ray sources when observed in
the optical and near-infrared (e.g. Barger et al. 2002). From these studies it is
apparent that the frequency and distribution of gas and dust plays a
major r\^ole in determining how we see these quasars.

Thus, any alternative method in which
these obscured quasars may be discovered would be extremely important.

In this letter we report the serendipitous discovery of a genuine
type-II quasar from large-volume integral-field observations with VIMOS
on the VLT.  

In section~\ref{sec:obs} we briefly summarise the
observing strategy and data reduction procedure and in
section~\ref{sec:results} we provide detailed information of the
type-II quasar including host galaxy properties and an estimate of its
black-hole mass. In section~\ref{sec:conc} we summarise our
conclusions and discuss how the discovery
of this object with integral-field observations may open up a new
window on the obscured AGN population.
We use $\Omega_{\rm M} = 0.3$, $\Omega_{\Lambda}= 0.7$ and $H_{\circ}
= 70$~km~s$^{-1}$~Mpc$^{-1}$ throughout, unless stated otherwise.

\section{Observations and data reduction}\label{sec:obs}

All of our observations were made on the nights of 29th April to the
2nd May 2003 with the Visible Multi-object Spectrograph integral-field
unit (VIMOS-IFU) on UT3 of the very large telescope (VLT). The IFU
consists of an array of $80 \times 80$ fibres coupled to
microlenses. Our observations were carried out in the low-resolution
mode with the LR-Blue grism giving a wavelength coverage of
3500\AA-7000\AA\, at 5.35\AA\, per pixel. We also used the
low-resolution spatial sampling which means that each fibre samples an
area of 0.67~arcsec resulting in a field-of-view per exposure of $54
\times 54$~arcsec$^{2}$.

The observations were centred on the powerful radio galaxy MRC0943-242
at $z = 2.92$ (R\"ottgering et al. 1995). The total exposure time was 9~hours, split into 18
exposures of 30~minutes which were dithered by 10~arcsec around the
central radio galaxy to enable accurate sky removal and cosmic-ray
rejection. All observations were taken in seeing of $< 0.8$~arcsec.

\begin{figure}
{\hbox to 0.48\textwidth{\null\null \epsfxsize=0.48\textwidth\epsfbox{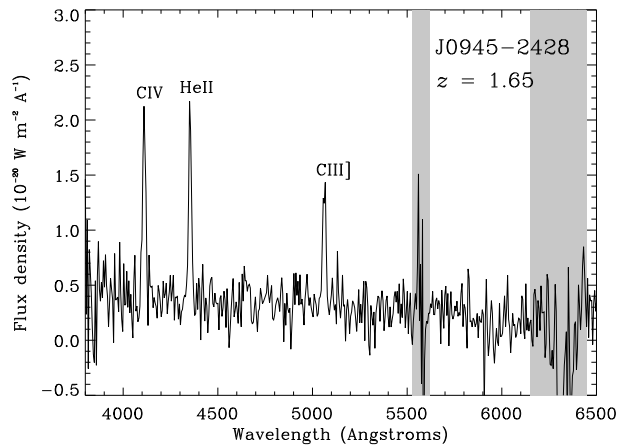}}}
{\caption{\label{fig:spec} The 1-D spectrum of the type-II quasar,
    J0945-2428 at $z = 1.65$. The spectrum was extracted over the
    whole galaxy (seven IFU fibres). The shaded regions show the wavelengths affected by sky-line emission. }}
\end{figure}

\begin{figure}
{\hbox to 0.48\textwidth{\null\null \epsfxsize=0.48\textwidth\epsfbox{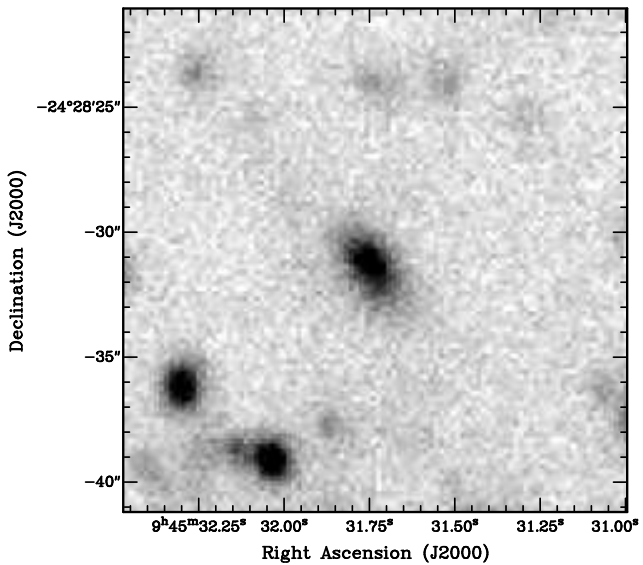}}}
{\caption{\label{fig:vband} V-band image of J0945-2428, clearly
    showing that the host galaxy is resolved.
}}
\end{figure}

We briefly discuss the data reduction here, although a more complete
description can be found in van Breukelen \&
Jarvis (in prep.).  To reduce the data we used the VIMOS Interactive
Pipeline and Graphical Interface (VIPGI) software (Scodeggio et
al. 2004) which takes the raw images and performs the usual bias
subtraction, flat-fielding, wavelength calibration and combining of
the individual frames.  For the IFU data the final dataset is combined
into a datacube of two-spatial dimensions and a third spectral
dimension. The spectrophotometric calibration is achieved via a script
in VIPGI which provides an accurate solution for the relative
flux-calibration between fibres, however, the absolute
flux-calibration is unreliable. Therefore, to overcome this we used a
calibrated spectrum of the central radio galaxy MRC0943-242
and summed the fibres in our IFU data cube along the slit direction,
and the V-band photometry of the type-II quasar, allowing us to bootstrap the spectrophotometry. The final flux
calibration is good to $\sim 20$~per cent.

The final datacube was used to find emission-line objects within the
volume probed with the IFU, from which we discovered the type-II
quasar discussed in this letter. A detailed investigation of the
number of Lyman-$\alpha$ emitters in the volume is deferred to a
subsequent paper (van Breukelen \& Jarvis in prep.).

\section{Results}\label{sec:results}
\begin{table*}
\begin{center}
\caption{\label{tab:emlines} Emission-line measurements for the type-II quasar J0945-2428. The uncertainties on the line-flux measurements are $\approx$30 per cent.}
\begin{tabular}{cccccc}\hline
Emission & $\lambda$ & $z$ & FWHM & Flux / $10^{-19}$ & $L / 10^{35}$ \\
line & (\AA) &  & (km~s$^{-1}$) & (W~m$^{-2}$) & (W) \\
\hline
CIV$\lambda$1549 & 4112 & 1.654 & $<1376$ & 3.5 & 6.3 \\
HeII$\lambda$1640 & 4355 &  1.655 & $<1127$ & 3.2 & 5.8 \\
CIII]$\lambda$1909 & 5065 & 1.653 & $<1167$ & 2.4 & 4.3 \\
\hline
\end{tabular}
\end{center}
\end{table*}

\subsection{A bona fide type-II quasar?}

Figure~\ref{fig:spec} shows the extracted 1-D spectrum of the type-II
quasar discovered in our datacube at a redshift of $z=1.65$ at
$\alpha=09^h45^m31.74^s$, $\delta=-24^{\circ}28^{\prime}31.3^{\prime\prime}$. Details
of the emission-line characteristics can be found in
table~\ref{tab:emlines}.  The type-II spectrum is very similar to that of
a powerful radio galaxy (see e.g. Jarvis et al. 2001a) with bright
narrow ($v < 1500$~km~s$^{-1}$) emission lines where the the CIV emission lines have luminosities of $10^{35}$~W$ < L_{\rm CIV} < 10^{36.5}$~W.  The host-galaxy
morphology is also resolved in the optical bands (see
figure~\ref{fig:vband}), implying
that the central engine is obscured along the line-of-sight, again similar to radio galaxies.

To confirm this as a genuine type-II quasar
and not just a radio galaxy we determine whether the source has any powerful radio emission. We use the 5~GHz radio map of Carilli et al. (1997) of the
field surrounding MRC0943-242\footnote{We also analysed the 1.4~GHz
  map from the VLA archive, but this was only sensitive to $\sim
  0.7$~mJy (3$\sigma$).}. At the position of the type-II quasar
there is no detectable radio emission down to a 3$\sigma$ sensitivity
of $S_{5 \rm GHz} = 0.15$~mJy. Given the redshift of the type-II quasar and
assuming a typical spectral index for the radio emission from the core
of a radio-quiet quasar of $\alpha = 0.7$ (Kukula et al. 1998),
this corresponds to a radio luminosity of $L_{5 \rm GHz} < 1.6 \times
10^{23}$~W~Hz$^{-1}$~sr$^{-1}$. This is a factor of 10 fainter than the traditional division between
radio-loud and radio-quiet quasars (e.g. Miller, Peacock \& Mead
1990). Thus, the object is radio-quiet.

Further, the emission-line ratios and equivalent widths are very
similar to radio galaxies, as opposed to quasars and Seyfert II
galaxies. The emission-line ratios of McCarthy (1993) show that for a
radio galaxy, the typical ratio of CIV/HeII= 1.1, and HeII/CIII] $\sim
$1.8, in J0945-2428 we find CIV/HeII= 1.1 and HeII/CIII]=1.3.  For
quasars and Seyfert II galaxies the same ratios from McCarthy (1993)
are (CIV/HeII)$_{\rm QSO}$ = 8.6 and (CIV/HeII)$_{\rm SyII}$ = 5.9 and
(HeII/CIII])$_{\rm QSO}$ = 0.26 and (HeII/CIII])$_{\rm SyII}$ =
0.37. Thus, the emission-line ratios are much more consistent with
those of powerful radio galaxies than other types of AGN.

The CIV/HeII ratio in J0945-2428 is however marginally smaller than
other type-II quasars discovered via their X-ray emission in the
literature. Stern et al. (2002) find that CXO~52 has a CIV/HeII
flux-ratio of $\approx 2$ and HeII/CIII] $\sim 0.8$, although the
CIV/HeII emission-line ratio of CDF-S 202 (Norman et al. 2002) is much
higher at CIV/HeII $\sim 3.5$, albeit still far from the ratio of a
typical quasar (e.g. Vanden Berk et al. 2001). This can possibly
explained by the the fact the CIV is a resonant line and as such some
of its flux may be absorbed along the line of sight, thus making it
appear fainter in our low-resolution observations. Another possible
effect is that carbon may deplete onto dust grains
(e.g. Villar-Mart\'\i n, Binette \& Fosbury 1996). Both of these factors
could easily reduce the CIV luminosity whereas the HeII luminosity
would remain roughly constant.

The only other emission lines that we would expect in our spectrum
would be OIII]$\lambda1663$ and CIII]$\lambda2326$. From our spectrum
we can set a 3$\sigma$ limit on the OIII]$\lambda1663$ emission line
of $< 4 \times 10^{34}$~W, at least a factor of 15 fainter than the
CIV line. This is again in line with observations of high-redshift
radio galaxies. The CIII]$\lambda2326$ line would lie at 6164\AA, and
is thus very close to the atmospheric emission highlighted in
Fig.~\ref{fig:spec}.

\subsection{The host galaxy of J0945-2428}

\begin{table}
\begin{center}
\caption{\label{tab:photo} Photometric properties of the type-II
  quasar. The B-band photometry was measured from a 75~minute exposure of this field with FORS2 on the VLT. The V- and I-band photometry was measured from 93~minute and 40~minute exposures respectively with LRIS on the Keck telescope.   }
\begin{tabular}{cl}\hline
Band & Magnitude \\
\hline
B & 23.00 $\pm 0.02$ \\
V & 22.88 $\pm 0.02$ \\
I & 22.2 $\pm 0.2$ \\
K & 20.5 $\pm 0.2$ \\
\hline
\end{tabular}
\end{center}
\end{table}

The type-II quasar has a continuum magnitude of $V = 22.8$,
which if we assume typical galaxy colours of an evolved elliptical
galaxy at $z = 1.65$ of $V - K \sim 5$ (Poggianti 1997) leads to $K
\sim 17.8$, typical of a radio galaxy at $z = 1.65$ (e.g. Jarvis et
al. 2001b; Willott et al. 2003). However, using archival NTT-SofI data
[programme ID: 70.A-0514(A)] on this field we find that the AGN has $K
= 20.5 \pm 0.2$. The $K-$band light samples the old stellar population
at this redshift and should therefore provide a good estimate of the
stellar mass of the host, but we find that the galaxy is much fainter
than the expected luminosity for a massive host galaxy. Indeed,
$K=20.5$ corresponds to $\sim 0.2~L^{\star}$ at $z = 1.65$ completely
at odds with the radio galaxy $K-z$ relation (e.g. Jarvis et
al. 2001b).

The broad-band colours (Table~\ref{tab:photo}) also show that the host
galaxy is extremely blue, with $B-I = 0.8$ and $I-K=1.7$, indicating
that the host may have a large amount of ongoing star
formation. Coupled with the faintness of the host in the near-infrared
this suggests that the host may be undergoing its first major
star-forming event and sub-millimetre observations would be able to
confirm this. 

There are also no obvious signs of merger activity, with the galaxy
profile exhibiting a disk-like morphology with a scale-length of $r =
9.2 \pm 0.7$~kpc, similar to the largest Seyfert galaxies
(e.g. Virani, De Robertis \& VanDalfsen 2000). However, it is worth
pointing out that we are sampling the UV-emitting stellar population
in our V-band (where we have the best signal-to-noise ratio), and that
the longer wavelength data sample the old stellar population
may not have such a morphology or scale-length, and further
observations will be need to investigate this.

\subsection{The mass of the black-hole in J0945-2428}\label{sec:mbh}

\begin{figure}
{\hbox to 0.48\textwidth{\null\null
\epsfxsize=0.48\textwidth\epsfbox{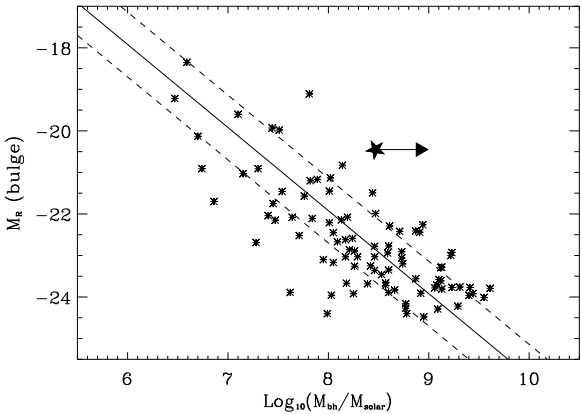}}}
{\caption{\label{fig:MbhMr} The position of J0945-2428 (filled star)
on bulge luminosity versus black-hole mass relation. We use the
back-hole mass derived from the strength of the UV emission lines as
detailed in section~\ref{sec:mbh}, this is a lower limit which assumes
the black hole is accreting at its Eddington limit. $M_{R}$ is
calculated from the fact that $K=20.5$ corresponds to 0.2$L^{\star}$
from the $K-$band luminosity function (Kochanek et al. 2001) and
$M_{R}^{\star} = -22.2$ (Lin et al. 1996). The solid line represents
the best fit relation from McLure \& Dunlop (2002) for inactive and
active galaxies at $z< 0.5$ and the small stars represent the objects
from their data set. All points are for $\Omega_{\rm M} =1$,
$\Omega_{\Lambda}=0$ and $H_{\circ} = 50$~km~s$^{-1}$~Mpc$^{-1}$ to
allow comparison with previous work.}}
\end{figure}

If we use the typical line ratios for radio galaxies then we are able
to estimate the strength of the [O{\footnotesize{II}}] emission lines
and gain an estimate of the bolometric luminosity, $L_{\rm Bol}$ of
the central engine. We use the relation given in Willott et al. (1999)
relating $L_{\rm Bol}$ to the luminosity of the
[O{\footnotesize{II}}]$\lambda3727$ line $L_{\rm [OII]}$, $L_{\rm Bol}
= 5 \times 10^{3} L_{\rm [OII]}$~W, along with the line ratios from
McCarthy (1993). Thus, assuming
C{\footnotesize{IV}}/[O{\footnotesize{II}}] = 1.0 means J0945-2428 has
a bolometric luminosity $L_{\rm Bol} \sim 3.2 \times 10^{39}$~W. If
the central engine is accreting at its Eddington limit, then this
corresponds to a black-hole mass of $\sim 3 \times 10^{8}$\Msolar.

Using the relation linking bolometric luminosity to the monochromatic
luminosity at 2500\AA ($L_{2500}$) from Elvis et al. (1994) and given
the relation between X-ray luminosity at 2~keV ($L_{2\rm keV}$), and
$L_{2500}$ from Risaliti, Elvis \& Gilli (2002) with an optical--X-ray
spectral index of $\alpha_{OX} = 1.6$, this provides a very rough
estimate of the expected X-ray flux at 2~keV. Combining these
relations shows that J0945-2428 would have an X-ray luminosity of
$L_{2\rm keV} \sim 10^{37}$~W. This is also consistent with taking the
typical line ratios found in radio galaxies and linking the X-ray
luminosity to the H$\beta$ luminosity (e.g. Ward et al. 1988).

These properties are typical of a powerful AGN at these redshifts and imply that
it is in the regime of black-hole mass where powerful radio
emission is able to be produced (e.g. McLure \& Jarvis 2004). 
However, such a massive black hole should be hosted by a massive
galaxy if the relation between black-hole mass and bulge luminosity
(Magorrian et al. 1998)
holds to high redshifts. From our $K-$band image we find that this is
not the case, which in turn implies that the black-hole mass -- bulge
luminosity correlation may break down (Fig.~\ref{fig:MbhMr}) in the early Universe. Thus
massive black holes may be in place before the host galaxy has fully
built up. However, this is highly tentative and obviously more observations are
needed of this object and other such quasars before such a statement
can be made in earnest.

\subsection{Obscuration toward the nucleus of J0945-2428}

To estimate the amount of obscuration toward the quasar nucleus we
adopt the Bolometric correction of $L_{\rm Bol} = 5.6 \times L_{2500}$ from Elvis et al. (1994), to calculate the continuum flux
expected in our optical spectrum. This leads to a monochromatic
luminosity at rest-frame 2500\AA\, of $L_{2500} = 5.3 \times
10^{38}$~W. From our extracted spectrum (Fig.~\ref{fig:spec}) we find
a monochromatic luminosity of $L_{2500} \sim 10^{34}$~W. This leads to
a visual extinction of A$_{\rm V} \gtsim 4$~mag toward the quasar
nucleus, which can easily explain the lack of a nuclear point source
in the $K-$band image. Again this is in line with studies of powerful
radio galaxies (e.g. Simpson et al. 1999).

 \subsection{The space density of obscured AGN}

\begin{figure}
{\hbox to 0.48\textwidth{\null\null \epsfxsize=0.48\textwidth\epsfbox{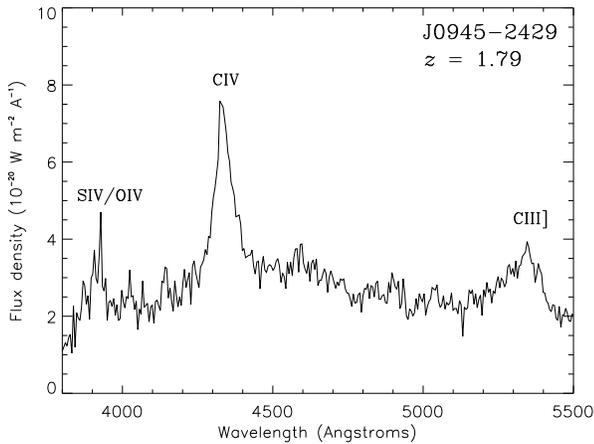}}}
{\caption{\label{fig:blqso} The 1-D spectrum of J0945-2429, the $z = 1.79$ unobscured quasar in our IFU volume. The spectrum was extracted over three IFU fibres.
}}
\end{figure}

Obviously with a single object it is very difficult to estimate the
space density of obscured AGN in the high-redshift Universe. However,
it is interesting to note that there are two other AGN in the volume
probed by our IFU observations, the powerful radio galaxy that we
targeted and a traditional broad-line quasar at $z = 1.79$ at
$\alpha=09^{h}45^{m}32.92^{s}$, $\delta=-24^{\circ}29^{\prime}09.1^{\prime\prime}$  (Fig.~\ref{fig:blqso}). Therefore, the best we can say is that obscured AGN are
consistent with being as abundant as their type-I counterparts, but
obviously a larger volume search with IFUs will be able to pin-down
this fraction in a highly efficient and unbiased manner.

\section{Conclusions}\label{sec:conc}

We have reported the first discovery of a genuine type-II quasar from
optical data alone. This was made possible by the unique capabilities
of the wide-field integral-field unit VIMOS on the VLT. 

J0945-2428 has narrow emission lines ($v <1500$~km~s$^{-1}$), a
resolved host galaxy and no detectable radio emission down to a flux-density
limit of $S_{5 \rm GHz} = 0.15$~mJy ($L_{5\rm GHz} = 1.6 \times
10^{23}$~W~Hz$^{-1}$~sr$^{-1}$) at $z = 1.65$, indicative of a bona fide
radio-quiet type-II quasar.

Using typical emission-line ratios of powerful radio galaxies and
assuming that it is accreting at its Eddington limit, we show that
J0945-2428 contains a supermassive black hole of mass $\sim 3 \times
10^{8}$\Msolar. However, from deep $K-$band imaging we find that the
host galaxy has a luminosity of $\sim 0.2~L^{\star}$, placing this
type-II quasar approximately $3\sigma$ away from the relation between
galaxy bulge luminosity and black-hole mass. This may be indicative of
the supermassive black hole in this object being in place before the
host galaxy is fully formed. This is also reinforced by the very blue
colour of the host galaxy, indicative of a large amount of ongoing
star formation activity. Although further observations of this source
and others are needed to confirm these results.

This serendipitous discovery of an obscured quasar highlights the way
in which wide-area integral-field units on large telescope could open up a
unique window on the Universe. VIMOS is currently the only instrument
that has the capability of large spectral coverage coupled with a
$\sim 1$~square arcminute field of view. However, future instruments,
such as the Multi-Unit Spectroscopic Explorer
(MUSE; http://clio.univ-lyon1.fr/MUSE/), will expand the initial work which is taking place with VIMOS.

Furthermore, obscured AGN are not the
only extragalactic sources which may be discovered with volumetric
surveys. Surveys with integral-field units will discover other objects such as bare quasars with unusual optical
spectra which precludes them from being identified as AGN in
colour-colour diagrams, Lyman-$\alpha$ emitters (van Breukelen \&
Jarvis in prep.) and possibly objects yet to be discovered.

\section*{ACKNOWLEDGEMENTS} 

We thank Chris Carilli for providing the 5~GHz map of
the field surrounding MRC0943-242, Wil van Breugel for the V-, R- and
I-band images and spectrum of MRC0943-242, Steve Rawlings for looking
at the 1.4~GHz map and Ross McLure for
supplying the data points shown in Fig.~3. We also thank the referee
for useful comments.
MJJ and RJW acknowledge the support of PPARC PDRAs and CVB would like to acknowledge the financial support from the ERASMUS program.

{}

\end{document}